\documentclass[letterpaper]{article}
\usepackage{amsmath}      
\usepackage{amssymb}
\usepackage{amsthm}
\usepackage{graphicx}
\usepackage{subcaption} 
\usepackage{natbib,alifeconf}  
\usepackage{url,hyperref,cleveref}
\usepackage{booktabs}
\hypersetup{ pdfborder={0 0 0}, } 

\title{Evolvable Chemotons: Toward the Integration of Autonomy and Evolution}

\author{
    Kuzuya Horibe$^{1,*}$,
    Daichi G. Suzuki$^{2,3,*}$ \\
    \mbox{}\\
    $^1$ RIKEN-CBS, JAPAN\\
    $^2$ University of Tsukuba, JAPAN\\
    $^3$ Corresponding author: suzuki.daichi.gp@u.tsukuba.ac.jp\\
    $^*$ These authors contributed equally to this work.
} 

\begin{document}

\maketitle

\begin{abstract}
    In this study, we provide a relatively simple simulation framework for constructing artificial life (ALife) with both autonomous and evolutionary aspects by extending G\'{a}nti's chemoton model. While the original chemoton incorporates metabolism, membrane, and genetic templates, it lacks a mechanism for phenotypic variation, preventing true evolutionary dynamics. To address this, we introduced a genotype–phenotype coupling by linking templates to a second autocatalytic cycle, enabling mutations to affect phenotype and be subject to selection. Using a genetic algorithm, we simulated populations of chemotons over generations. Results showed that chemotons without access to the new cycle remained in a stable but complexity-limited regime, while lineages acquiring the additional metabolic set evolved longer templates. These findings demonstrate that even simple replicator systems can achieve primitive evolvability, highlighting structural thresholds and rare innovations as key drivers. Our framework provides a tractable model for exploring autonomy and evolution in ALife.
    
\end{abstract}

Submission type: \textbf{Late Breaking Abstracts}\\

Code available at: \url{https://github.com/KazuyaHoribe/chemoton}

\section{Introduction}

It is widely recognized that \textit{autonomy} and \textit{evolution} are essential factors for life \citep{Ruiz-Mirazo2004-ec}. In artificial life (ALife) research, however, these two factors have tended to be studied separately, with a few exceptional studies. 

One of the most pioneering (and less discussed in the field of ALife) examples is the chemoton model of cells \citep{Ganti1975-ke, Ganti2003-qm}. A chemoton consists of three subsystems: metabolic cycles, membrane boundaries, and genetic templates. This triadic characterization of life has been widely accepted \citep{Szathmary2005-ea, Rasmussen2016-ef, Sole2009-xo, Nurse2020-en}. The first two (metabolism and boundaries) represent the autonomous or autopoietic aspect of life \citep[see][for the comparison]{Luisi2006-lz}, while the last (templates) indicates the potential to incorporate genetic inheritance. However, this G\'{a}ntian model does not fully accommodate the evolutionary aspect of life because it focuses primarily on a single individual chemoton, not a lineage or population. 

For the integration of the autonomy and evolution, here we offer an extended chemoton model encompassing these two aspects. The classical chemoton offers autopoietic subsystems (i.e., the metabolism, membrane, and template), but its template lacks phenotypic effect, so genetic changes cannot generate novelties subject to selection. To overcome this disadvantage, we add a genotype-phenotype coupling by letting templates have access to a second autocatalytic loop, and we examined the populational dynamics of these chemotons through simulational evolution.

\begin{figure}[t]
    \centering
    \includegraphics[width=5.5cm]{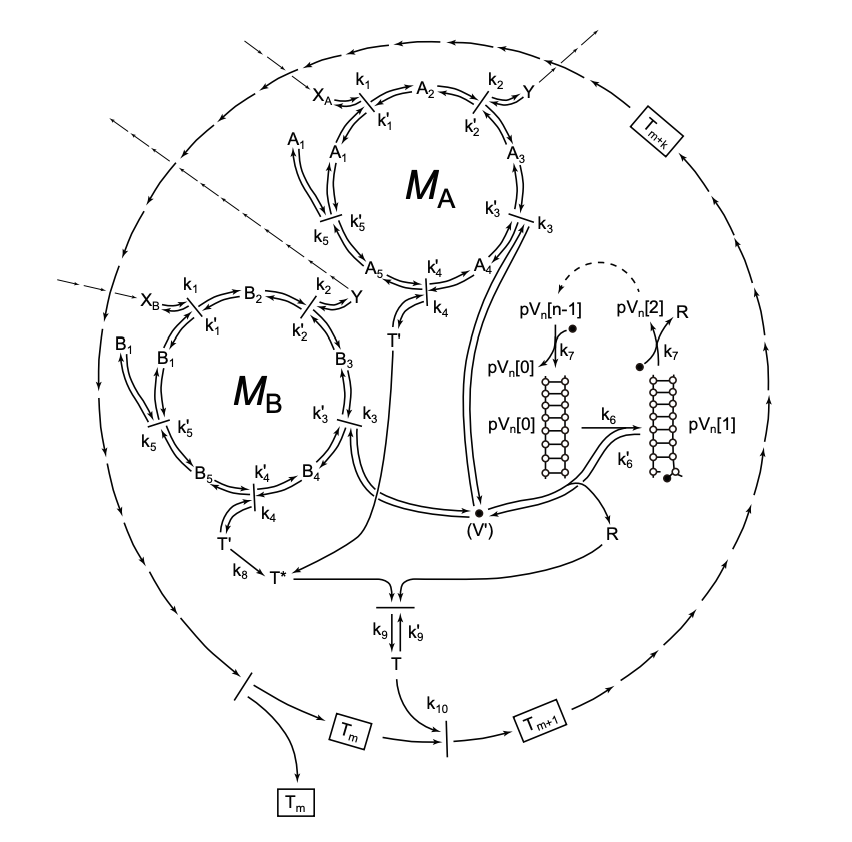}
    \caption{
        \textbf{The ``evolvable chemoton'' model.} A chemoton with two metabolic cycles, based on Fernando and Di Paolo's (\citeyear{fernando2004chemoton}) model (refer to this paper for abbreviations).
    }
    \label{fig:models}
\end{figure}

\section{Methods}
We first prepared an initial population with $P=10$ chemotons only with a gene set for metabolism A ($M_\mathrm{A}$), and then ran the genetic algorithm with the following growth, selection, and mutagenesis phases for 50 generations. The biochemical reactions for metabolism, membrane growth, and template replication were described by ordinary differential equation (ODE) system, acccording to previous studies \citep{csendes1984simulation, fernando2004chemoton}. All initial chemotons were set to have $N\approx25$ templates under the normal distribution. {\it Growth phase}: ODEs were solved through LSODA until 1.0 simulation time, during which a default chemoton can divide 9 times; chemotons that reached a surface-area threshold were allowed to divide into two daughter chemotons, copying their $N$ and the gene set status coded in their templates. {\it Selection phase}: the grown population was randomly selected to $P$ individuals---thus trajectory that divided more had a chance to be selected. {\it Mutagenesis phase}: every survivor or selected chemoton received a Gaussian perturbation ($\sigma=2$) to $N$ for mutation. If a chemoton obtained a new template with $N\ge40$ and it lacked the gene set for metabolism B ($M_\mathrm{B}$), it was allowed to acquire the gene set for metabolism B with a probability $p = 0.3$. If the chemoton with gene sets for both metabolism A and B obtained a new template with $N$ below 40, the chemoton was forced to lose the gene set for metabolism B. Chemotons with $N<20$ templates were discarded as lethal. 

All state variables were initialized as follows. The series of metabolite intermediates $[A_1]$–$[A_5]$ began at [1.0, 1.8, 1.9, 1.7, 10.0], and $[B_1]$–$[B_5]$ were all zero (inactive). Intracellular monomer concentration $V$ started at 40.0, and the initial template abundance $pV_0$ was set to 0.01.  The residue pool $R$ was 0.5. Membrane precursor states $T_P$, $T_S$, and $T$ were initialized to 17, 14, and 0, respectively.  Surface‐area and volume proxy variables $S$ and $Q$ both began at 1.0.  The smoothing variable \texttt{tmpl\_len} was set equal to $N$, and the Boolean metabolism‐B flag \texttt{has\_met\_B} was initialized to \texttt{false}.

\section{Results}
\begin{figure}[t!]
    \centering
    \includegraphics[width=\linewidth]{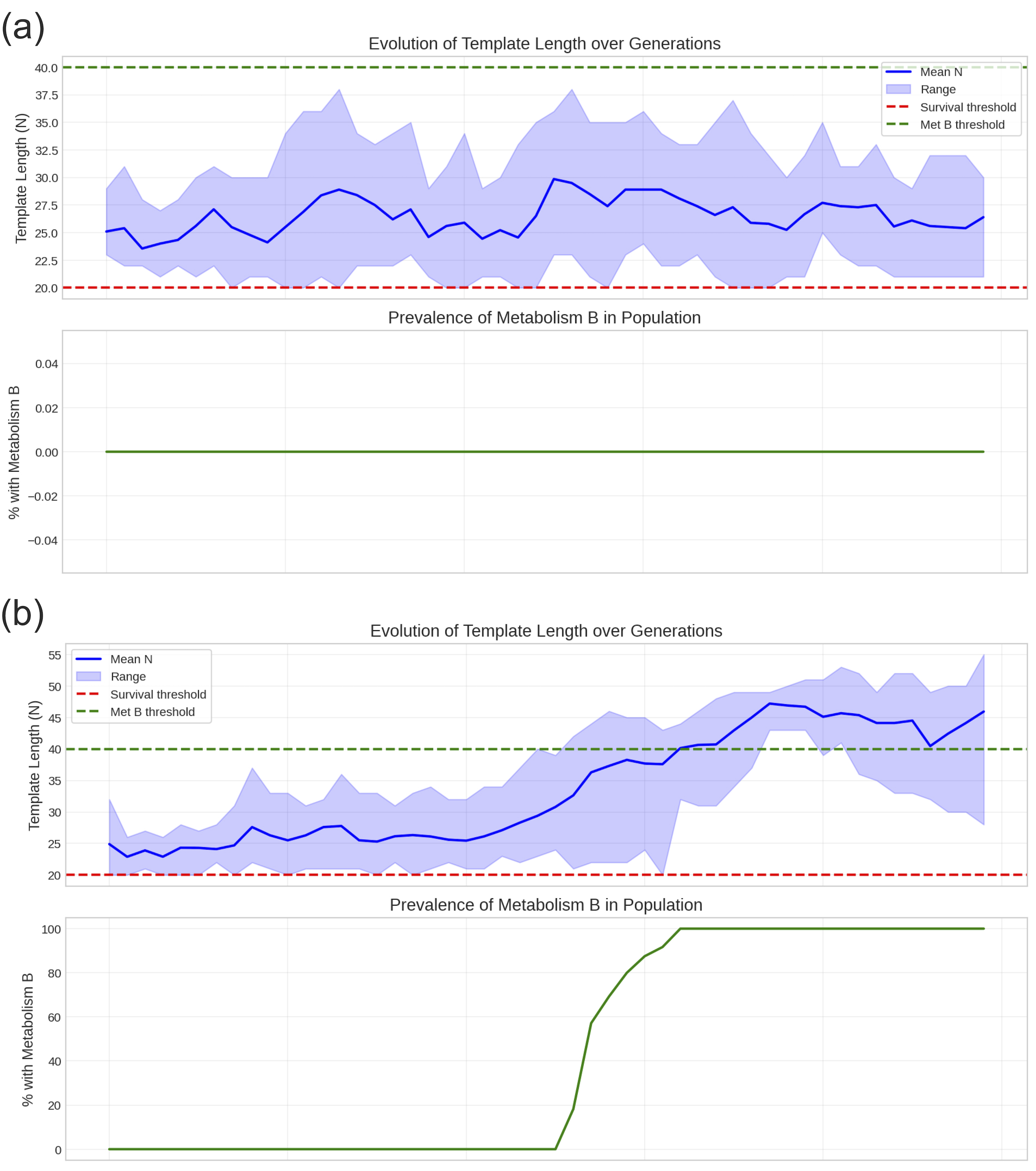}
    \caption{\textbf{Outcomes for the evolvable chemoton model.} (a) B-lacking and (b) B-acquired trajectories. For each trajectory, the mean template length N with range (above; blue lines with light blue bands) and the prevalence of metabolism B in the population (below) are shown. Horizontal red and green dashed lines mark the survival threshold (N=20) and the acquisition threshold for metabolism B (N=40), respectively.}
    \label{fig:results}
\end{figure}

As shown in Figure \ref{fig:results}a, the B-lacking trajectory keep $N$ oscillating between 22–32, never breaching the gate $N=40$. In contrast, B-acquired trajectory shows a pivotal event: once any cell surpasses $N=40$ and wins thdraw of probability $p$ (In this simulation, $p=0.3$), B carriers divide faster, pushing the mean $N$ into the 40–50 zone and drithe prevalence of B to 100\% in 5 generations (Figure~\ref{fig:results}b).

These results highlight that evolutionary success requires both structural preconditions (such as sufficiently long templates) and rare beneficial innovations (exemplified by metabolism B). Remarkably, even under realistic mutational and probabilistic constraints, adaptive evolution remains accessible in a substantial fraction of runs. To conclude, our evolvable chemoton model demonstrates that simple replicator systems can robustly evolve toward greater complexity and productivity.

\section{Discussion}
The simulation results shown here suggest that our model manifests a primitive sort of evolvability and adaptation, as evolution is typically defined as temporal changes in the genetic frequency within a gene pool; although shorter templates facilitate a high cell division rate, some populations acquire new gene sets for an additional metabolic cycle by retaining longer templates. In addition, our results well illustrate the genomic evolution of organisms, as it is suggested that the increase of genome size is accompanied by the increase of organismal complexity \citep{Sharov2006}.

With the present model, we provide a relatively simple simulation framework for constructing ALife with both autonomous and evolutionary aspects. This model can be used for various advanced analyses. For example, it should be worth to examine the responce to environmental changes of nutrient supply. Also, our model can be sophisticated to embody a more realistic coding system. One way for this to be achieved is by arranging a series of genes that code for respective enzymes involved in the biochemical reactions in the chemoton model. 

\section{Acknowledgements}
This work is supported by JSPS Grants-in-Aid for Scientific Research (Grant Numbers JP24H01538 for D.G.S. and JP24K20859 for K.H.).

\footnotesize
\bibliographystyle{apalike}
\bibliography{EvolvableChemoton} 

\end{document}